\input harvmac
\input graphicx
%
%
%
%
\ifx\includegraphics\UnDeFiNeD\message{(NO graphicx.tex, FIGURES WILL BE IGNORED)}
\def\figin#1{\vskip2in}
\else\message{(FIGURES WILL BE INCLUDED)}\def\figin#1{#1}
\fi
\def\Fig#1{Fig.~\the\figno\xdef#1{Fig.~\the\figno}\global\advance\figno
 by1}
%
%
%
%
\def\Ifig#1#2#3#4{
\goodbreak\midinsert
\figin{\centerline{
\includegraphics[width=#4truein]{#3}}}
\narrower\narrower\noindent{\footnotefont
{\bf #1:}  #2\par}
\endinsert
}
%
%
\font\ticp=cmcsc10

\def\calo{{\cal O}}

\def\mpl{{M_{p}}}
\def\mst{M_{st}}
\def\lst{l_{st}}

\def\mthsu{\mathsurround=0pt  }
\def\leftrightarrowfill{$\mthsu \mathord\leftarrow\mkern-6mu\cleaders
  \hbox{$\mkern-2mu \mathord- \mkern-2mu$}\hfill
  \mkern-6mu\mathord\rightarrow$}
 \def\overleftrightarrow#1{\vbox{\ialign{##\crcr\leftrightarrowfill\crcr\noalign{\kern-1pt\nointerlineskip}$\hfil\displaystyle{#1}\hfil$\crcr}}}
\overfullrule=0pt
%
%
\lref\ACVone{
  D.~Amati, M.~Ciafaloni and G.~Veneziano,
  ``Superstring Collisions At Planckian Energies,''
  Phys.\ Lett.\ B {\bf 197}, 81 (1987).
}
\lref\ACVtwo{
  D.~Amati, M.~Ciafaloni and G.~Veneziano,
  ``Classical And Quantum Gravity Effects From Planckian Energy Superstring
  Collisions,''
  Int.\ J.\ Mod.\ Phys.\ A {\bf 3}, 1615 (1988).
}
\lref\AmatiTN{
  D.~Amati, M.~Ciafaloni and G.~Veneziano,
 ``Can Space-Time Be Probed Below The String Size?,''
  Phys.\ Lett.\ B {\bf 216}, 41 (1989).
}
\lref\Bousrev{
  R.~Bousso,
  ``The holographic principle,''
  Rev.\ Mod.\ Phys.\  {\bf 74}, 825 (2002)
  [arXiv:hep-th/0203101].
}
\lref\tHooholo{
  G.~'t Hooft,
  ``Dimensional reduction in quantum gravity,''
  arXiv:gr-qc/9310026.
  }
\lref\Sussholo{
  L.~Susskind,
  ``The World as a hologram,''
  J.\ Math.\ Phys.\  {\bf 36}, 6377 (1995)
  [arXiv:hep-th/9409089].
}
\lref\Bous{
  R.~Bousso,
  ``A Covariant Entropy Conjecture,''
  JHEP {\bf 9907}, 004 (1999)
  [arXiv:hep-th/9905177]\semi
  ``Holography in general space-times,''
  JHEP {\bf 9906}, 028 (1999)
  [arXiv:hep-th/9906022].
}
\lref\SGinfo{S.~B.~Giddings,
  ``Quantum mechanics of black holes,''
  arXiv:hep-th/9412138\semi
  ``The Black hole information paradox,''
  arXiv:hep-th/9508151.
}
\lref\Astroinfo{
  A.~Strominger,
  ``Les Houches lectures on black holes,''
  arXiv:hep-th/9501071.
}
\lref\LPSTU{
  D.~A.~Lowe, J.~Polchinski, L.~Susskind, L.~Thorlacius and J.~Uglum,
  ``Black hole complementarity versus locality,''
  Phys.\ Rev.\ D {\bf 52}, 6997 (1995)
  [arXiv:hep-th/9506138].
}
\lref\GiLitwo{
  S.~B.~Giddings and M.~Lippert,
  ``The information paradox and the locality bound,''
  Phys.\ Rev.\ D {\bf 69}, 124019 (2004)
  [arXiv:hep-th/0402073].
}
\lref\GiLione{
  S.~B.~Giddings and M.~Lippert,
  ``Precursors, black holes, and a locality bound,''
  Phys.\ Rev.\ D {\bf 65}, 024006 (2002)
  [arXiv:hep-th/0103231].
}
\lref\tHooftRB{
  G.~'t Hooft,
  ``The Black hole horizon as a quantum surface,''
  Phys.\ Scripta {\bf T36}, 247 (1991).
}
\lref\SusskindIF{
  L.~Susskind, L.~Thorlacius and J.~Uglum,
  ``The Stretched horizon and black hole complementarity,''
  Phys.\ Rev.\ D {\bf 48}, 3743 (1993)
  [arXiv:hep-th/9306069].
}
\lref\tHooftRE{
  G.~'t Hooft,
  ``On The Quantum Structure Of A Black Hole,''
  Nucl.\ Phys.\ B {\bf 256}, 727 (1985).
}
\lref\tHooftFR{
  G.~'t Hooft,
  ``The Black Hole Interpretation Of String Theory,''
  Nucl.\ Phys.\ B {\bf 335}, 138 (1990).
}
\lref\Mald{
  J.~M.~Maldacena,
  ``The large N limit of superconformal field theories and supergravity,''
  Adv.\ Theor.\ Math.\ Phys.\  {\bf 2}, 231 (1998)
  [Int.\ J.\ Theor.\ Phys.\  {\bf 38}, 1113 (1999)]
  [arXiv:hep-th/9711200].
}
\lref\HoMa{
  G.~T.~Horowitz and J.~Maldacena,
  ``The black hole final state,''
  JHEP {\bf 0402}, 008 (2004)
  [arXiv:hep-th/0310281].
}
\lref\CaMa{
  C.~G.~.~Callan and J.~M.~Maldacena,
  ``D-brane Approach to Black Hole Quantum Mechanics,''
  Nucl.\ Phys.\ B {\bf 472}, 591 (1996)
  [arXiv:hep-th/9602043].
}
\lref\Hawkunc{
  S.~W.~Hawking,
  ``Breakdown Of Predictability In Gravitational Collapse,''
  Phys.\ Rev.\ D {\bf 14}, 2460 (1976).
}
\lref\DMW{
  J.~R.~David, G.~Mandal and S.~R.~Wadia,
  ``Microscopic formulation of black holes in string theory,''
  Phys.\ Rept.\  {\bf 369}, 549 (2002)
  [arXiv:hep-th/0203048].
}
\lref\Peet{
  A.~W.~Peet,
  ``TASI lectures on black holes in string theory,''
  arXiv:hep-th/0008241.
}
\lref\StVa{
  A.~Strominger and C.~Vafa,
  ``Microscopic Origin of the Bekenstein-Hawking Entropy,''
  Phys.\ Lett.\ B {\bf 379}, 99 (1996)
  [arXiv:hep-th/9601029].
}
\lref\Hawknew{
  S.~W.~Hawking,
  ``Information loss in black holes,''
  Phys.\ Rev.\ D {\bf 72}, 084013 (2005)
  [arXiv:hep-th/0507171].
}
\lref\GMH{
  S.~B.~Giddings, D.~Marolf and J.~B.~Hartle,
  ``Observables in effective gravity,''
  arXiv:hep-th/0512200.
}
\lref\Lowe{
  D.~A.~Lowe,
 ``Causal properties of string field theory,''
  Phys.\ Lett.\ B {\bf 326}, 223 (1994)
  [arXiv:hep-th/9312107].
}
\lref\DaVe{
  T.~Damour and G.~Veneziano,
  ``Self-gravitating fundamental strings and black holes,''
  Nucl.\ Phys.\ B {\bf 568}, 93 (2000)
  [arXiv:hep-th/9907030].
}
\lref\HoPotwo{
  G.~T.~Horowitz and J.~Polchinski,
  ``Self gravitating fundamental strings,''
  Phys.\ Rev.\ D {\bf 57}, 2557 (1998)
  [arXiv:hep-th/9707170].
}
\lref\HoPoone{
  G.~T.~Horowitz and J.~Polchinski,
  ``A correspondence principle for black holes and strings,''
  Phys.\ Rev.\ D {\bf 55}, 6189 (1997)
  [arXiv:hep-th/9612146].
}
\lref\LSU{
  D.~A.~Lowe, L.~Susskind and J.~Uglum,
  ``Information spreading in interacting string field theory,''
  Phys.\ Lett.\ B {\bf 327}, 226 (1994)
  [arXiv:hep-th/9402136].
}
\lref\Ginew{S.B. Giddings, ``Black hole information, unitarity, and nonlocality," hep-th/0605196.}
\lref\GrMe{
  D.~J.~Gross and P.~F.~Mende,
  ``String Theory Beyond The Planck Scale,''
  Nucl.\ Phys.\ B {\bf 303}, 407 (1988).
}
\lref\MeOo{
  P.~F.~Mende and H.~Ooguri,
  ``Borel Summation Of String Theory For Planck Scale Scattering,''
  Nucl.\ Phys.\ B {\bf 339}, 641 (1990).
}
\lref\Sund{
  B.~Sundborg,
  ``High-Energy Asymptotics: The One Loop String Amplitude And Resummation,''
  Nucl.\ Phys.\ B {\bf 306}, 545 (1988).
}
\lref\GIZ{
  E.~Gava, R.~Iengo and C.~J.~Zhu,
  ``Quantum Gravity Corrections From Superstring Theory,''
  Nucl.\ Phys.\ B {\bf 323}, 585 (1989).
}
\lref\ABC{
  M.~Ademollo, A.~Bellini and M.~Ciafaloni,
  ``Superstring Regge Amplitudes And Emission Vertices,''
  Phys.\ Lett.\ B {\bf 223}, 318 (1989).
}
\lref\thoo{
  G.~'t Hooft,
  ``Graviton Dominance In Ultrahigh-Energy Scattering,''
  Phys.\ Lett.\ B {\bf 198}, 61 (1987).
}
\lref\MuSo{
  I.~J.~Muzinich and M.~Soldate,
  ``High-Energy Unitarity Of Gravitation And Strings,''
  Phys.\ Rev.\ D {\bf 37}, 359 (1988).
}
\lref\Ver{
  H.~L.~Verlinde and E.~P.~Verlinde,
  ``Scattering at Planckian energies,''
  Nucl.\ Phys.\ B {\bf 371}, 246 (1992)
  [arXiv:hep-th/9110017].
}
\lref\Venunpub{G. Veneziano, unpublished.}
\lref\DiEm{
  S.~Dimopoulos and R.~Emparan,
  ``String balls at the LHC and beyond,''
  Phys.\ Lett.\ B {\bf 526}, 393 (2002)
  [arXiv:hep-ph/0108060].
}
\lref\SaSk{
  P.~Salomonson and B.~S.~Skagerstam,
  ``On Superdense Superstring Gases: A Heretic String Model Approach,''
  Nucl.\ Phys.\ B {\bf 268}, 349 (1986);
``Strings At Finite Temperature,''
  PhysicaA {\bf 158}, 499 (1989)
}
\lref\MiTu{
  D.~Mitchell and N.~Turok,
  ``Statistical Mechanics Of Cosmic Strings,''
  Phys.\ Rev.\ Lett.\  {\bf 58}, 1577 (1987); 
   ``Statistical Properties Of Cosmic Strings,''
  Nucl.\ Phys.\ B {\bf 294}, 1138 (1987).
}
\lref\Damo{
  T.~Damour,
  ``The entropy of black holes: A primer,''
  arXiv:hep-th/0401160.
}
\lref\PoSt{
  J.~Polchinski and M.~J.~Strassler,
  ``Hard scattering and gauge/string duality,''
  Phys.\ Rev.\ Lett.\  {\bf 88}, 031601 (2002)
  [arXiv:hep-th/0109174].
}
\lref\SGFroiss{
  S.~B.~Giddings,
  ``High energy QCD scattering, the shape of gravity on an IR brane, and  the
  Froissart bound,''
  Phys.\ Rev.\ D {\bf 67}, 126001 (2003)
  [arXiv:hep-th/0203004].
}
\lref\Penrose{R. Penrose, unpublished (1974).}
\lref\EaGi{
  D.~M.~Eardley and S.~B.~Giddings,
``Classical black hole production in high-energy collisions,''
  Phys.\ Rev.\ D {\bf 66}, 044011 (2002)
  [arXiv:gr-qc/0201034].
}
\lref\BaFi{
  T.~Banks and W.~Fischler,
  ``A model for high energy scattering in quantum gravity,''
  arXiv:hep-th/9906038.
}
\lref\Hsu{
  S.~D.~H.~Hsu,
 ``Quantum production of black holes,''
  Phys.\ Lett.\ B {\bf 555}, 92 (2003)
  [arXiv:hep-ph/0203154].
}
  \lref\GiRy{S.~B.~Giddings and V.~S.~Rychkov,
  ``Black holes from colliding wavepackets,''
  Phys.\ Rev.\ D {\bf 70}, 104026 (2004)
  [arXiv:hep-th/0409131].
}
\lref\GiTh{
  S.~B.~Giddings and S.~D.~Thomas,
  ``High energy colliders as black hole factories: The end of short  distance
  physics,''
  Phys.\ Rev.\ D {\bf 65}, 056010 (2002)
  [arXiv:hep-ph/0106219].
}
\lref\DiLa{
  S.~Dimopoulos and G.~Landsberg,
  ``Black holes at the LHC,''
  Phys.\ Rev.\ Lett.\  {\bf 87}, 161602 (2001)
  [arXiv:hep-ph/0106295].
}
\lref\Thorne{
  K.~S.~Thorne,
  ``Nonspherical Gravitational Collapse: A Short Review,'' in J. R. Klauder, {\sl Magic without magic} (San Francisco 1972).
}
\lref\YoNa{
  H.~Yoshino and Y.~Nambu,
  ``Black hole formation in the grazing collision of high-energy particles,''
  Phys.\ Rev.\ D {\bf 67}, 024009 (2003)
  [arXiv:gr-qc/0209003].
}
\lref\Porev{
  J.~Polchinski,
  ``String theory and black hole complementarity,''
  arXiv:hep-th/9507094.
}
\lref\BaFiholo{
  T.~Banks and W.~Fischler,
  ``Holographic cosmology 3.0,''
  Phys.\ Scripta {\bf T117}, 56 (2005)
  [arXiv:hep-th/0310288].
}
\lref\CKN{
  A.~G.~Cohen, D.~B.~Kaplan and A.~E.~Nelson,
  ``Effective field theory, black holes, and the cosmological constant,''
  Phys.\ Rev.\ Lett.\  {\bf 82}, 4971 (1999)
  [arXiv:hep-th/9803132].
}
\lref\VenezianoER{
  G.~Veneziano,
  ``String-theoretic unitary S-matrix at the threshold of black-hole
  production,''
  JHEP {\bf 0411}, 001 (2004)
  [arXiv:hep-th/0410166].
}
\lref\AiSe{
  P.~C.~Aichelburg and R.~U.~Sexl,
  ``On The Gravitational Field Of A Massless Particle,''
  Gen.\ Rel.\ Grav.\  {\bf 2}, 303 (1971).
}
\lref\PiazzaGU{
  F.~Piazza,
  ``Quantum degrees of freedom of a region of spacetime,''
  arXiv:hep-th/0511285.
}
\lref\BHMR{
  S.~B.~Giddings,
  ``Black holes and massive remnants,''
  Phys.\ Rev.\ D {\bf 46}, 1347 (1992)
  [arXiv:hep-th/9203059].
}
\Title{\vbox{\baselineskip12pt
\hbox{hep-th/0604072}
}}
{\vbox{\centerline{Locality in quantum gravity and string theory}
}}
\centerline{{\ticp Steven B. Giddings}\footnote{$^\ast$}
{Email address:
giddings@physics.ucsb.edu} }
\centerline{ \sl Department of Physics}
\centerline{\sl University of California}
\centerline{\sl Santa Barbara, CA 93106-9530}
\bigskip
\centerline{\bf Abstract}

Breakdown of local physics in string theory at distances longer than the string scale is investigated.  Such nonlocality would be expected to be visible in ultrahigh-energy scattering.  The results of various approaches to such scattering are collected and examined.  No evidence is found for nonlocality from strings whose length grows linearly with the energy.  However, local quantum field theory does apparently fail at scales determined by gravitational physics,  particularly strong gravitational dynamics.  This amplifies locality bound arguments that such failure of locality is a fundamental aspect of physics.  This kind of nonlocality could be a central element of a possible loophole in the argument for information loss in black holes.

\Date{}

\newsec{Introduction}

Einstein's theory of general relativity and local quantum field theory have existed in profound theoretical conflict for scores of years.  There are various manifestations of this conflict, but it particularly comes into focus in the black hole information paradox, which emerged from Hawking's argument\refs{\Hawkunc} that black holes destroy information and thus violate quantum mechanics -- for reviews see \refs{\SGinfo,\Astroinfo}.
There has been a longstanding hope that the existence of such a sharp statement of the clash between gravity and field theory would serve as a useful guide to unearthing the principles needed to mesh these theoretical frameworks, and this has motivated much work on the subject.

The information paradox pushes us to abandon one of the cherished principles of physics.  A critical idea of 't Hooft's was that the paradox should be resolved in favor of quantum mechanics, but at the sacrifice of locality.  This idea advanced through important work of Susskind and others to find a more concrete formulation in the hypothesized {\it holographic principle}\refs{\tHooholo\Sussholo\Bous-\Bousrev}, together with the  {\it principle of black hole complementarity}\refs{\tHooftRE\tHooftFR\tHooftRB-\SusskindIF}, which are believed to summarize key features of black hole dynamics.  In particular, the holographic principle states that the number of microstates of a black hole is proportional to its surface area, rather than its volume, as local quantum field theory  would predict.  And the principle of black hole complementarity states that different observers in a black hole spacetime can see different physics, but it is not possible to directly compare their experience and thus produce a contradiction.

There are two things notably lacking from such a picture.  One is an explanation of what precisely is wrong with Hawking's original argument\refs{\Hawkunc} for information loss.  And, more generally, we lack a full description of the underlying nonlocal microphysics responsible for such a picture.  There have been a number of hints from string theory, such as counting of microstates of certain black holes in certain regimes\refs{\StVa,\CaMa} (for reviews see \refs{\Peet,\DMW}) and the AdS/CFT correspondence\refs{\Mald}, but as yet string theory has been woefully inadequate to the task of penetrating the veil of mysteries surrounding black holes that are far from extremality. 

In the meantime, Hawking has rejected his original calculation\Hawkunc.  Although he has suggested a new approach to the problem\refs{\Hawknew}, it is hard to discern from this approach what is actually wrong with the original calculation.

Without a full understanding of the microphysics of quantum gravity, we can take a more modest approach to these problems by attempting to answer the linked questions of 1) under what conditions is locality expected to fail 2) what physics is responsible, and 3) does this address the information paradox.
This approach attempts to draw a boundary around the domain where we should seek a more fundamental, and nonlocal, microphysics. 

Local field theory obeys basic axioms, such as existence of field operators, and the statement that local operators commute at spacelike separations.  A more fundamental theory should reduce to quantum field theory as a limit, and produce fields, local observables, and locality in this limit.\foot{For discussion of how local observables can approximately arise from a diffeomorphism-invariant framework, see \refs{\GMH}.}  So, part of the question is where this reduction fails, and a proposal for an answer is the {\it locality bounds} of refs.~\refs{\GiLione,\GiLitwo}.

One obvious possible source of nonlocality arises from the extended nature of strings.  There have been some preliminary investigations of nonlocality intrinsic to string theory\refs{\Lowe,\LSU}, and this kind of nonlocality has been suggested to play a direct role in resolving the information paradox\refs{\LPSTU}.  Specifically, \LPSTU\ consider commutators of string field operators, and argue that 
{\it long strings} can contribute to commutators outside the light cone, and in particular play a role in resolving the information paradox.  

One such argument roughly suggests that if one considers a commutator of two operators with significant overlap with states with total high energy $E$, the fact that a string of this energy can stretch over a distance 
\eqn\SLB{L\sim E/\mst^2\ ,}
where $\mst=1/\lst$ is the string mass scale,
 indicates that there will be nonlocal contributions to the commutator.

Another, more generic, potential source of nonlocality was argued in \refs{\GiLione,\GiLitwo} to arise from strong gravitational effects.  Roughly, if two nearby local operators with sufficiently large combined energy act on the vacuum, their energy strongly deforms the spacetime and there is no longer a justification for local field theory to apply.  An approximate criterion for this proposed breakdown is that the separation between the operators be less than the Schwarzschild radius corresponding to their center-of-mass energy, 
\eqn\GLB{L\roughly<R_S(E)\ .}

Either of these possible effects would only arise at high energies, above the string or Planck scales.  Here it is important to stress a key assumption of this work, namely that Lorentz invariance is an exact feature of the fundamental theory.  This means that one can describe an ultra-planckian energy particle by viewing a low-energy particle from a highly boosted frame.  At the semiclassical level, a good description of the corresponding gravitational field is known: it is the Aichelburg-Sexl metric\AiSe.   Exact Lorentz invariance appears to be a feature of string theory, but is not assumed in some other approaches to quantum gravity.  It is the viewpoint of this paper that arbitrarily large boosts can be applied at the kinematical level to individual particles or collections of particles. Of course, at the dynamical level,  large {\it relative} boosts can have enormous effects, for example by producing strong interactions when a highly-boosted particle interacts with a low-energy particle.  But such exact Lorentz invariance gives us an apparently sharp starting point for considering ultra-planckian collisions, since we can imagine independently preparing two widely separated individual ultrahigh-energy particles, and then allowing them to collide.  

For spacetime dimension $D>4$, and for $D=4$ with weak string coupling, a string ``locality bound" would become important before the gravitational one:  $R_S(E)<E/\mst^2$.  An important question is to determine which of these possible effects actually occurs, and in particular which could be relevant to resolving the black hole information paradox.

Indeed, the relative role of these two possible sources of nonlocal physics is potentially important for another question:  do black holes form in high-energy collisions?  The argument that they do rests on semiclassical arguments within local quantum field theory, but 
if local field theory were to break down due to string effects at a scale \SLB\ in a collision with energy $E$, this would undermine application of local field theory to argue for creation of black holes on the shorter scale $R_S(E)$.  

This paper will turn this question around.  Specifically, one can study ultra-high energy scattering in string theory, and look for possible modifications due to effects that correct local field theory.  Drawing from direct studies of high-energy scattering\refs{\ACVone\ACVtwo\AmatiTN\GrMe-\MeOo}, related discussions of the black hole/string correspondence principle\refs{\HoPoone\HoPotwo-\DaVe}, and high-energy scattering behavior in the AdS/CFT correspondence\SGFroiss, one finds no evidence in favor, and significant evidence against, nonlocal string effects presenting themselves on scales $L\sim E$.  On the other hand, there is of course strong evidence for breakdown of local field theory on the gravitational scale $L\sim R_S(E)$, and,  following \refs{\GiLitwo,\Ginew}, this physics has been suggested to provide a  loophole in Hawking's argument for information loss in black holes.

The next section will summarize some aspects of such high-energy scattering, and discuss the question of how one could explain the absence of string effects on scales $L\sim E$.  Section three then gives a more general discussion of breakdown of locality.  Following the above reasoning, it is argued that an important role is played by {\it strong gravitational effects} in line with the gravitational locality bound.   In the process, a more general gravitational locality bound, for $N$-particle systems, is formulated.  This section also further discusses the proposals\refs{\GiLitwo,\Ginew}
that such gravitational nonlocality produces a loophole in arguments like Hawking's for information loss; in short, the nonlocal physics plausibly  invalidates the assumption that the Hilbert spaces inside and outside the black hole are independent.  The discussion also potentially sheds light on the role of the ``final state" proposal of Horowitz and Maldacena\refs{\HoMa}, and on the relative roles of the locality bound and the holographic principle.

\newsec{Scattering at super-planckian energies}

Consider a Gedanken experiment with an accelerator that collides electrons and protons at an arbitrarily high tunable center-of-mass energy $E$.  As the energy is tuned through $\sim 1\ GeV$, one enters the regime of Bjorken scaling; scattering is interpreted in terms of a field theory of increasingly weakly interacting quarks and gluons rather than one of free protons.  At increasing energies, perhaps other thresholds of compositeness are passed, introducing more fundamental field-theoretic degrees of freedom.  The scattering behavior as a function of $E$ and momentum transfer $q$, or equivalently $E$ and impact parameter $b$, would reveal a great deal about the quantum field theory dynamics.  But for what regimes of these parameters would one expect {\it any} local field theory description to {\it fail}? One of course expects breakdown of spacetime when $E>\mpl$, on planckian distance scales, and similar effects from strings, at string scales.  But our focus will be on the question of whether, for sufficiently high energy, locality is violated at much longer distances.

The two obvious proposals for physics violating local field theory are string dynamics, and strong gravitational dynamics.  

String dynamics has been anticipated to lead to a breakdown of locality by virtue of the extended nature of strings; see {\it e.g.} \refs{\Lowe,\LSU}.  The threshold for such behavior should be the string scale, $E\sim\mst$, at which one expects locality to fail on distance scales $\sim\lst$.  However, at higher energies one might guess that nonlocality is manifest on longer scales, since the available energy could produce a string of length $\sim E/\mst^2$.  Specifically, the resultant nonlocality, if present, would be expected to influence scattering at impact parameters 
\eqn\bstr{b\sim E/\mst^2\ ,}
 and therefore lead to failure of local quantum field theory on these scales.  

On the other hand, strong gravitational dynamics is also expected to lead to failure of local quantum field theory, when energies are sufficient to strongly deform spacetime\refs{\GiLione,\GiLitwo}.  Consider the general case of a $D$-dimensional spacetime.  Gravity becomes strong when a given energy is concentrated in a region comparable to its Schwarzschild radius, so in the collision context, for impact parameters
\eqn\bgrav{b\roughly< R_S(E)={1\over\mpl} \left({E\over \mpl}\right)^{1\over {D-3}}\ .}

An important question regards the relative role of string and gravitational dynamics.  Indeed, the na\"\i ve expectation that string dynamics produces breakdown of local field theory at impact parameter \bstr\ would suggest a very limited role for gravitational dynamics.  If one considers collisions at a fixed energy $E\gg \mst$, and progressively lower impact parameter, for $D>4$ such string dynamics would become relevant first.  The presence of string nonlocality of the colliding particles would not allow one to conclude that the strong gravitational regime could ever be reached, since such nonlocality could prevent localization of the energy on the Schwarzschild scale \bgrav.  Even for $D=4$, a similar conclusion would be reached for weak string coupling, since there $\mpl>\mst$.
In short, presence of string nonlocality on scales \bstr\ could prevent the conclusion that black holes form in a controllable approximation in high-energy collisions.

\subsec{High-energy string scattering}

Some understanding of high-energy scattering in string theory is thus needed to clarify the relative roles of string and gravitational effects.  Significant work has been done on this subject, 
in classic papers by Amati, Ciafaloni, Veneziano\refs{\ACVone\ACVtwo-\AmatiTN} (ACV), and by Gross, Mende, and Ooguri\refs{\GrMe,\MeOo}.  While there are still open questions, these works, together with other more recent explorations, suggest the outlines of a picture.

\Ifig{\Fig\figphase}{A ``phase" diagram for high energy scattering in string theory.  At a given large energy $E$, by decreasing the impact parameter $b$ one first encounters a possible regime of ``long strings," then the impact parameter corresponding to the fixed-$t$ regime, followed by the regime where ``diffractive scattering" (tidal string excitation) becomes relevant, and finally the regime of strong gravity, below the Schwarzschild radius, $R_S(E)$.  The validity of the analysis of \refs{\GrMe,\MeOo} is limited to the region $E<E_{GMO}$.}{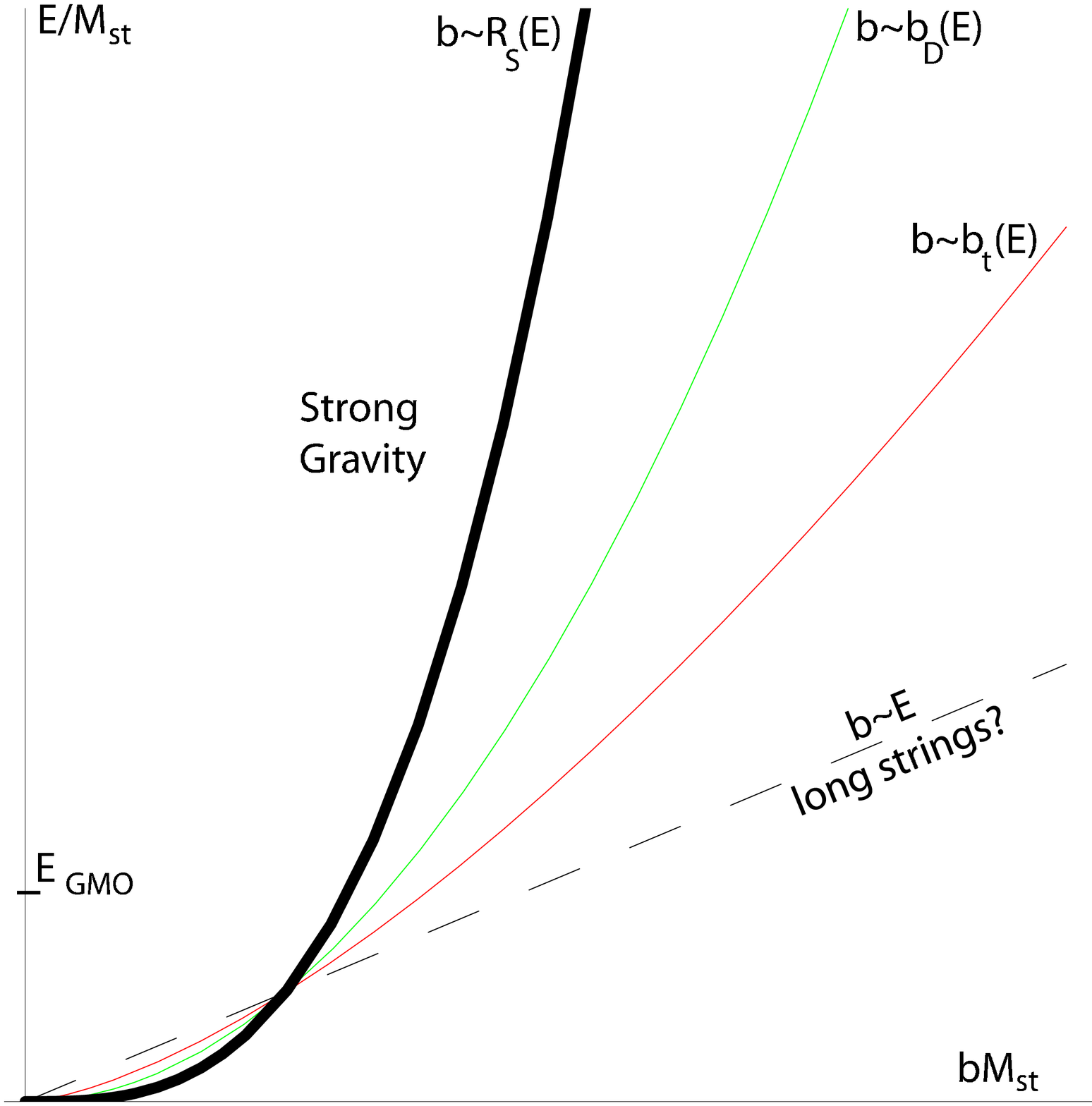}{5.5}

Beginning with \refs{\ACVone,\ACVtwo}, ACV investigated high-energy string scattering at large energy, using Regge-Gribov techniques to resum an infinite class of string diagrams.  Their initial analysis treated large $E$ and fixed $t$, but \ACVtwo\ argues that the fixed $t$ condition can be relaxed, and that their results are in fact valid for impact parameters $b\roughly>max(R_S(E),1/\mst)$. ACV's approach has also been checked by explicit one-loop\refs{\Sund} and two-loop\refs{\GIZ,\ABC} string calculations. 
A summary of their picture is as follows; these results can be better understood by consulting \figphase.

For energies $E\gg \mpl$ and impact parameters
\eqn\bdiff{b\roughly> b_D(E) \sim {1\over \mpl} \left({E\over \mst}\right)^{2\over D-2}\ ,}
ACV find scattering in agreement with the long-distance Coulomb-like result expected from gravity; specifically, agreement was found with treatments of high energy gravitational scattering by `t Hooft\refs{\thoo}, Muzinich and Soldate\refs{\MuSo}, and Verlinde and Verlinde\refs{\Ver}.  In fact, ACV  argue\refs{\ACVtwo,\Venunpub} that corrections to the eikonal result are consistent with scattering not in a Schwarzschild metric, but rather in the Aichelburg-Sexl metric, which is expected to be the appropriate metric for describing a high-energy collision.  

At impact parameters $b\sim b_D(E)$, part of the amplitude is lost to ``diffractive scattering;" there is a non-negligible amplitude to excite states of the colliding strings.  The value of $b_D$ and other features like the average excitation mass fit well with a simple picture (further described in section 2.3):  the string has internal dynamics, and at impact parameter $\sim b_D$ tidal forces from the gravitational field excite its vibrational modes during the collision.

This analysis does not reveal any other large corrections to the expected gravitational scattering for $b\roughly> b_I\sim \log E$.  For $R_S(E)\roughly>1/\mst$, there are large corrections  to the leading eikonal result, at scales $b\sim R_S(E)$, and in particular strong absorption of the amplitude.  The presence of such large corrections is of course expected, and corresponds to the onset of strong gravitational dynamics such as black hole formation.  

Thus there is no evidence for modifications to this dynamics on scales $b\sim E$, which would represent the kind of nonlocality arising from a single long string, as discussed in \LPSTU.
One might ask why one does not see nonlocal effects due to strings on these scales in the scattering picture of ACV, and whether this picture could possibly have missed such effects.  We turn to this question next.  It is nonetheless conceivable that strings play a role in scattering and locality below the scale set by $b_D(E)$.  We discuss that dynamics, and its relation to black hole formation, in section 2.3.

\subsec{No long strings?}

\Ifig{\Fig\dual}{Creation of a virtual long string is dual to long-distance exchange of a short string.}{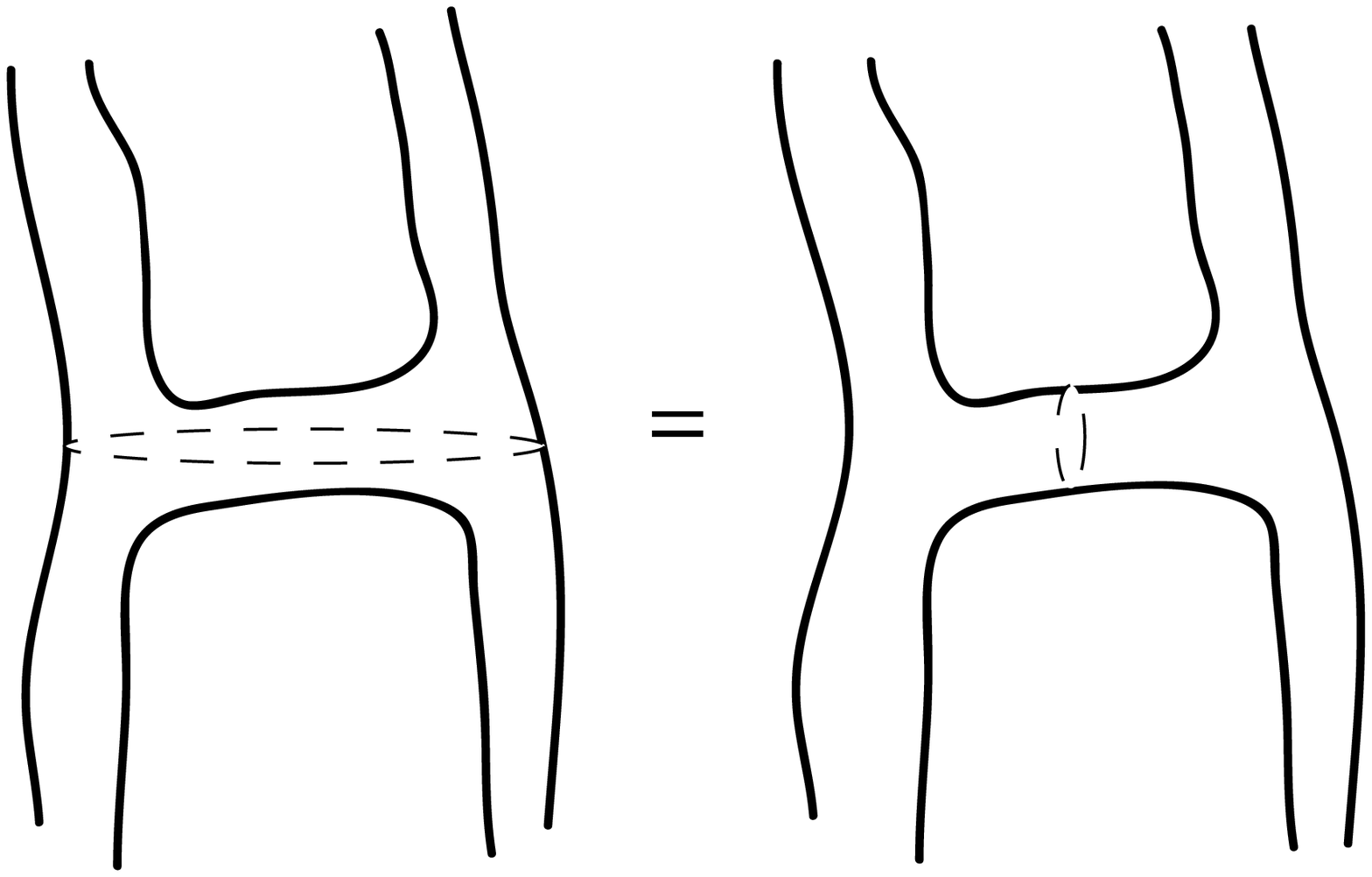}{5.5}

Note that a piece of our explanation arises from duality.    
Creation of a long virtual closed string, is, as pictured in \dual, dual to exchange of a short closed string, corresponding to graviton exchange.  The interactions producing this exchange are nonlocal on the string scale, but not necessarily on larger scales.

One might also expect to be able to create real long-string states at large $E$, and indeed tree amplitudes, for example in two-to-two string scattering, exhibit such poles.  However, they are presumably difficult to create due to small form factors, and 
moreover interactions are very important for such states. Indeed they are expected to be broad (rapidly decay), and  furthermore the tree level amplitude for their creation unitarizes at modest energies\refs{\Venunpub,\DiEm}, $E\sim 1/g_s$.  A string with size $R\sim E$ is expected to be both very atypical and unstable, and to rapidly break up into pieces or otherwise decay.  This happens due to processes -- splitting and joining of pieces of the string -- that can be described as approximately local, at least down to the string distance scale, and is plausibly dual to multiple graviton effects analogous to \dual.

Indeed, the atypicality of such strings follows from consideration of generic string configurations with total (large) energy $E$. If one ignores interactions, these have argued to be well-described by random walk models in \refs{\SaSk,\MiTu}, and correspondingly have typical size $R\sim\sqrt E$.  We could inquire whether string modifications to local field theory are expected at {\it this} distance scale.  Certainly the calculations of \refs{\ACVone\ACVtwo} suggest not, as one sees from \bdiff\ for $D>6$.  Once again one expects that interactions are very important and these are not stable states.  These (approximately local) interactions will lead to both fragmentation into short strings and gravitational collapse of the string configuration.  In considering the former, note that the typical entropy of a gas of gravitons of total energy $E$ in a volume of size $R\sim \sqrt E$ is much higher than that of the string:  $S_{grav}\sim E^{3D/2(D+1)}$ as compared to $S_{st} \sim E$.  Gravitational collapse has also been argued to be an important process in destabilizing such string configurations, and plays a direct role on the black-hole/string correspondence principle\refs{\HoPoone\HoPotwo-\DaVe,\Damo}.  There are thus good arguments that the typical localized high energy state in string theory is a black hole, above the energy where $R_S\sim\lst$.  
Both fragmentation and gravitational collapse thus could play a role in explaining why there aren't new scattering contributions on scales $b\sim\sqrt E$ from string states.

It is also of interest to consider the relation with results of Gross and Mende\refs{\GrMe}, who study high-energy scattering at fixed {\it angle}, and thus large $t$.  Ref.~\GrMe\ found, at a given order in the loop expansion, saddle point Riemann surfaces that give dominant contributions to the fixed-angle amplitude.  The contribution at $N$-loop order, for given scattering angle $\phi$, behaves as 
\eqn\gmest{{\cal A}_N\propto g_s^N e^{-{E^2f(\phi)/ N}}\ ,}
and thus for large energies, large order dominates.  A Borel resummation of the resulting series was performed by Mende and Ooguri\MeOo.  This shows, roughly, that the amplitude is dominated by contributions with $N\sim E$, which one sees directly in \gmest.  The Riemann surface in question corresponds to an intermediate string wound on itself $N$ times.  Thus, in this picture the increasing energy does not go into creating a string stretched over size $E$ or even $\sqrt E$; the spatial extent of the string stays of order $1/\mst$.  This is in harmony with the viewpoint discussed above that string effects only become manifest on string scales.  It is not, however, a direct check of this statement in the regime of interest to us, since it only a statement about what processes contribute to fixed angle scattering, and moreover, since Mende and Ooguri's analysis breaks down at relatively low energies, $E_{GMO}\sim [-\ln g_s^2]^{3/2}$.

\subsec{Tidal string excitation and black hole formation}

Next consider the question of the physics associated to the scale $b_D(E)$, eq.~\bdiff.  We can straightforwardly see that this is associated with excitation of the colliding strings by tidal forces, as follows.

\Ifig{\Fig\shock}{Deflection of geodesics in an Aichelburg-Sexl geometry.  Shown is the side view of the gravitational shock wave (thick vertical line) of a particle with an ultrahigh right-moving velocity.  Geodesics that pass through the shock are focussed inward in its wake.}{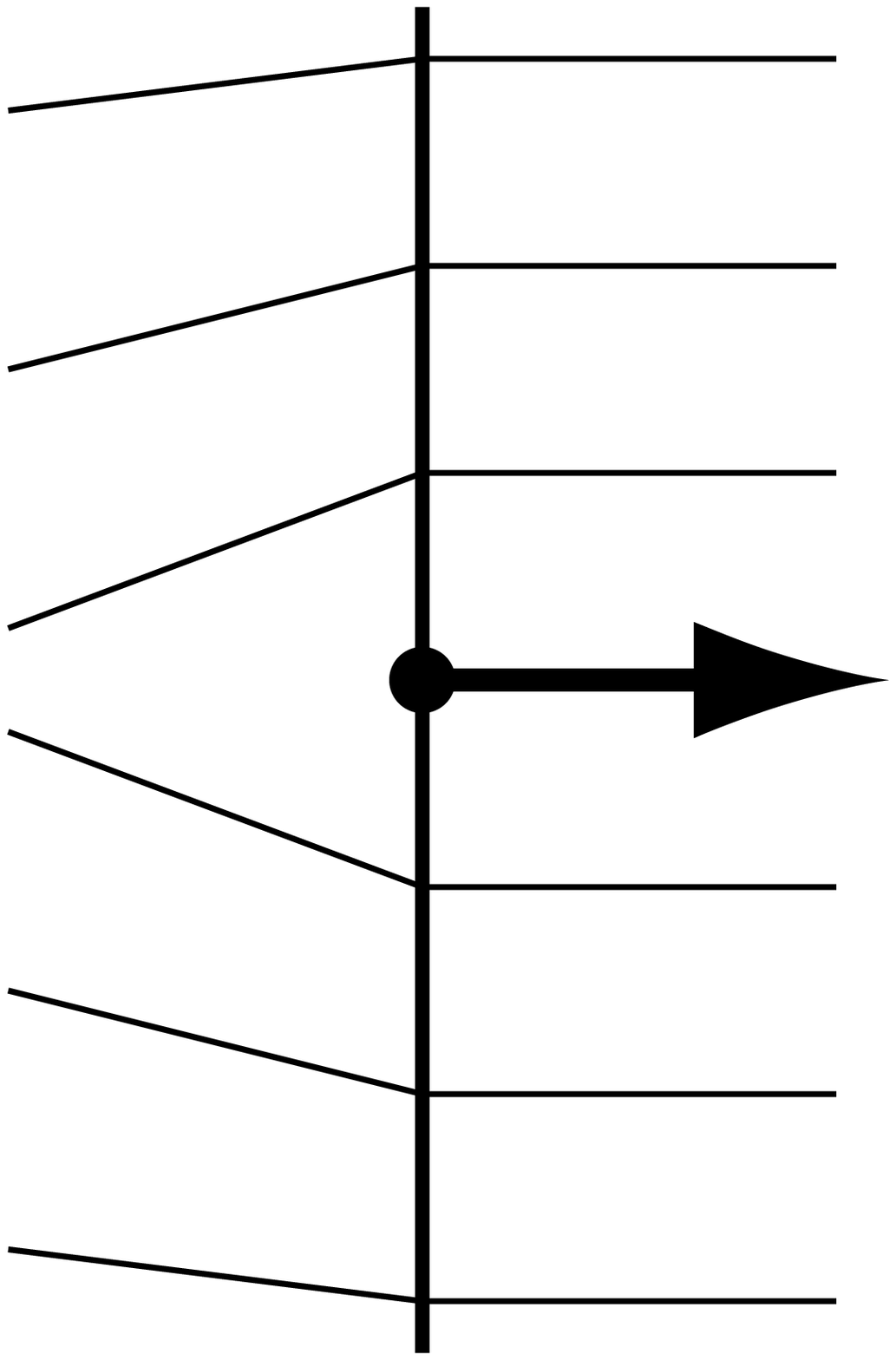}{2.0}

Consider the collision in a ``lab" frame where one of the strings has moderate energy $E_0$.  In this frame, the second string has energy
\eqn\elab{E_l\sim {E^2\over E_0}\ .}
We will estimate the effect of the gravitational field of the high-energy string on the low-energy string.  To do so, note that at long distances this gravitational field should be given by the Aichelburg-Sexl metric\refs{\AiSe}, generalized to the relevant dimension:
\eqn\aisemet{ds^2 = -du dv + dx^{i2} + \Phi(x^i) \delta(u) du^2\ .}
Here $u=t-y$ and $v=t+y$ are light-cone coordinates, and $x^i$ are the transverse coordinates.  The high-energy string travels along the curve $u=x^i=0$.  The function $\Phi$ is essentially a gravitational potential
\eqn\Phidef{\Phi(x^i) = {k\over \mpl^{D-2} }{E_l\over \rho^{D-4}}}
 in the transverse dimensions, with radius $\rho^2=x^{i2}$, and $k$ a constant.  (In $D=4$, one instead has a log.)  As a geodesic crosses the surface $u=0$ from right to left, it experiences both a shift and its tangent is redirected; see \shock.  For example, a particle initially moving in the $-x$ direction gets a kick in the $x^i$ direction when it crosses the shock.  This force is described by the Christoffel symbol,
 \eqn\Christsy{\Gamma^i_{uu} \propto \partial^i \Phi \delta (u)\ .}

 Since the low-energy string is an extended object of size $\sim \lst$, the resulting force has a tidal differential, producing an  acceleration differential between the two transverse extremities of the string:
 \eqn\tidalf{{d \Delta p^i\over dt} \sim \Gamma^i_{uu}(b+\lst) -\Gamma^i_{uu}(b) \propto {E_l \delta(u) \over b^{D-2}}\ ,}
 where $p^i$ is the momentum of an element of string.
 The string excitation amplitude should be proportional to this relative kick, in string units, integrated over $u$, which serves to ``pluck" the string.  The condition for this amplitude to become of order unity reproduces the energy dependence in $b_D$, \bdiff, which was derived directly  from an approximation to string scattering in \refs{\ACVone,\ACVtwo}.  
 
 This does by itself portend any nonlocality -- one simply has gravitational scattering of composite objects that can become excited.  However, when the string excitation stretches the strings so that their size is comparable to their separation, one might imagine such an effect could be possible.  An estimate of when this happens comes from noting that constant tension implies the stretch is proportional to $\Delta p$, so becomes comparable to $b$ for impact parameter $b_T(E)$ which solves
 \eqn\tidal{b_T\sim {E^2\over b_T^{D-2}}\ .}

Without a deeper understanding we cannot rule out some important effects, and in particular some nonlocality, at the distance scale $b_T(E)$.  However, we can ask whether it is sufficient to undermine arguments for relevance of strong gravitational physics at shorter scales.  

The results of \refs{\ACVone,\ACVtwo} again suggest no:  ACV see a modest decrease in the elastic cross section, but still evidence for a breakdown of the eikonal description at $b\sim R_S(E)$.  They suggest that the interpretation of this behavior is that a ``black disk" becomes grey.

Moreover, evidence for gravitational dominance at high energy and large impact parameter also comes from the AdS/CFT correspondence.  Consider a version of this correspondence where conformal symmetry is broken by some dynamics that can be effectively described as truncating AdS space in the infrared, as described for example in \refs{\PoSt}.  One can then investigate the correspondence between high-energy scattering in the gauge theory and  string theory in the (truncated) AdS space.  In gauge theory, the known upper bound on the total scattering cross-section is the Froissart bound, $\sigma\propto \ln^2 E$.  A corresponding energy dependence is found on the string theory side\refs{\SGFroiss}, and arises from bulk configurations where gravity is becoming strongly coupled  -- in the AdS context, $R_S\propto \ln E$.  This thus fits with the gravitational picture.

Yet one more argument for the relevance of a strong gravitational domain is recent work of Veneziano\refs{\VenezianoER}, who argues that evidence of black hole behavior can be seen in string scattering as one approaches the  expected black hole region from the domain below the production energy threshhold.

A way to understand the persistent importance of strong gravity follows from a different approximation in which one  can study some aspects of high energy collisions at impact parameters $b\roughly<R_S(E)$:  working in a semiclassical expansion, with expansion  parameter $\mpl/E$, about the classical geometry that arises\refs{\Penrose,\EaGi} from the collision of the gravitational shock-waves of the two strings.  These are again well approximated by the Aichelburg-Sexl solution, and their collision can be shown to form a trapped surface and hence a black hole\refs{\EaGi}.\foot{For further discussion of the semiclassical approach to high-energy scattering, see \refs{\BaFi\Hsu-\GiRy}.}   This picture suggests both arguments of causality and dynamics for why tidal excitation doesn't  undermine black hole formation. 

The first argument rests on causality.  The colliding strings are of ``size" $\calo(\lst)$, and the trapped surface of \refs{\Penrose,\EaGi} is of size $R_S$ and moreover forms {\it before} the two Aichelburg-Sexl shock waves intersect.  Thus by the time the strings can experience a tidal force, they are inside the trapped region.

The second, dynamical, argument follows from study of \shock.  The tidal forces on the string are radial and tend to pull parts of one string differentially towards the center of the other's Aichelburg-Sexl metric.  However, this will act to stretch the string only until it reaches that center; on the other side of that center the force is directed in the opposite direction.  Once the collision of the shockwaves takes place, one expects the subsequent gravitational field to exhibit the same behavior.  So the gravitational field is expected to act to concentrate the strings into a region behind the horizon.

In short, the possibility of interesting dynamics at the tidal scale $b_T$ cannot be ruled out, and it is not inconceivable it could contribute to an intrinsically stringy form of nonlocality.\foot{It could also be interesting from the phenomenological viewpoint, in TeV-scale gravity scenarios, as a correction to high-energy scattering\refs{\GiTh,\DiLa}.}  But there are also apparently good arguments that strong gravitational dynamics sets in at $b\sim R_S(E)$, and here there is very good reason to expect that local field theory breaks down from strongly coupled gravitational effects.

\subsec{High-energy string scattering -- summary}

These arguments suggest a picture of high-energy scattering, in the ultraplanckian regime, described as follows.

\item{1.}  There is no evidence in high-energy string scattering for modification of local quantum field theory on scales that would arise from long strings,\foot{While branes haven't been explicitly accounted for, a related hypothesis is that they likewise don't lead to long-distance nonlocalities of this sort.} at impact parameters $b\sim E$ or even $b\sim\sqrt E$.  Moreover, there is considerable evidence that there are no such modifications. 

\item{2.} At the scale $b_D(E)\sim E^{2/(D-2)}$ (eq.~\bdiff), excitation of strings by the tidal forces resulting from the long-range gravitational field becomes relevant.  However, the dynamics is still apparently local.

\item{3.} At the scale $b_T(E)\sim E^{2/(D-1)}$ (eq.~\tidal), tidal forces can stretch the strings to a size comparable to the impact parameter.  This could lead to new effects, and our arguments haven't ruled out the possible that this dynamics is nonlocal.

\item{4.} Evidence and arguments exist that, despite such tidal effects, strong gravity should set in, and in particular black holes should form, at scales $b\sim R_S(E)$.  Here there is good reason for local quantum field theory to fail.  While  the semiclassical approximation correctly describes some aspects of this process,  we will argue in more detail that at a more basic level local field theory breaks down here due to strong gravitational effects.
\vskip.1in

This section has compiled evidence in favor of this picture, but the question of high-energy scattering behavior remains a very interesting one.  It is  plausible that in the high-energy context in string theory, modifications to local field theory only become evident at scales set by gravitational physics, rather than physics intrinsic to string theory; this would suggest that string theory doesn't necessarily play a  direct role in such a breakdown of local field theory.  However, in light of the tidal dynamics that has been described, it is also not inconceivable that nonlocalities intrinsically of combined string/gravity origin also have a role to play.

\newsec{Locality and its breakdown}

\subsec{Strings vs. gravity}

While scattering can exhibit evidence for locality or its breakdown, one might more typically phrase the question as an off-shell one.  Our study of 
scattering does not appear to exhibit breakdown of local field theory as a result of string effects at distance scales $b\sim E/\mst^2$,
but does appear to exhibit such breakdown due to gravitational effects at scales $b\sim R_S(E)$.  What general consequences does this have for the domain of validity of local field theory?

Locality in field theory is often stated as the condition that gauge-invariant local operators commute outside the lightcone,
\eqn\localcom{[\calo(x),\calo(y)]=0\quad {\rm for} \quad (x-y)^2>0\ .}
However, such a statement is ill-defined in the context of a theory with gravity.  First, the local operator $\calo(x)$ contains modes of all momenta up to infinity; one would thus expect a large backreaction, particularly for the combined operator in \localcom.  Thus suggests working with wavepackets,  
\eqn\wavedef{\calo[f] = \int d^D x f(x) \calo(x)\ .}
If $f$ and $g$ are two functions of compact and spacelike-separated support, the Wightman axioms include the statement that 
\eqn\wight{\left[\calo[f],\calo[g]\right]=0\ .}
Alternatively, one might consider gaussians or other wavepackets with strong falloff, in which case equivalently the expression \wight\ would vanish up to terms arising from small tails of the wavepackets\refs{\GiLione}.

Even use of wavepackets does not suffice to consistently define nearly-local observables in the context of a gravitational theory.  Observables must be gauge invariant, and the gauge invariance of 
the low-energy limit of gravity includes diffeomorphism invariance.  For this reason, ref.~\refs{\GMH} argues that the appropriate generalization of local observables in a theory with gravity are {\it relational} observables, which approximately reduce to local observables in appropriate states.  A toy example is furnished by the $\psi^2\phi$ model, with two scalar fields $\psi$ and $\phi$.  In this case, a diffeomorphism invariant expression is
\eqn\psiphi{\calo_{\psi^2\phi} = \int d^Dx \sqrt{-g} \psi^2(x)\phi(x)\ , }
and in a state corresponding to appropriately arranged incident wavepackets of the $\psi$ field, the expectation value of a product of such operators approximately reduces to a vacuum correlator of operators of the form
\eqn\phiop{\int d^Dx\sqrt{-g} f(x) \phi(x)\ ,}
analogous to \wavedef. 

In this sense, the ``generalized observables" of the form \psiphi\ approximately reduce to field theory observables, but we anticipate breakdown of the approximation which yields this result in certain limits.  These are suggested to represent {\it fundamental} limitations on recovery of local field theory\refs{\GiLione,\GiLitwo,\GMH}.

We could attempt to examine locality in this approach.  Specifically, consider working about a flat background and with operators $\phi_{x,p}$, corresponding to the special case of an operator of the form \phiop\ in which a particle is created at position $\approx x$ with momentum $\approx p$, with an appropriate gaussian spread in each.    For example, taking $|0\rangle$ to be the (gravitationally  dressed) $\phi$-vacuum, we could under what circumstances an expression like
\eqn\twophi{\phi_{x,p} \phi_{y,q}|0\rangle}
ceases to obey axioms corresponding to the Wightman axioms of local quantum field theory.

A na\"\i ve expectation is that, for $\phi_{x,p}$ derived as an appropriate limit in string theory, the expression \twophi\ would exhibit nonlocal behavior when
\eqn\stringloc{|x-y|\roughly< {|p+q|\over \mst^2}\ }
corresponding to exceeding the threshold center of mass energy required to create a string stretching between $x$ and $y$.
Ref.~\refs{\LPSTU} argued that a similar kind of string nonlocality could be the nonlocality responsible for resolving the black hole information paradox.  (For prior studies of locality in string theory, see \refs{\Lowe,\LSU}.)  

However, the previous section explained that there is no evidence of stringy nonlocal behavior in scattering on corresponding scales; to reiterate, the results of \refs{\ACVone,\ACVtwo} are consistent with a description based on long-distance gravitational scattering at distances far shorter than given by \stringloc.

\subsec{Gravitational locality bounds}

On the other hand, and taking the scattering discussion as a guide, a local field theory description of \twophi\ does appear to break down when
\eqn\schwloc{|x-y|\roughly> R_S(|p+q|)\ ,}
is violated, corresponding to strong gravity/quantum black hole formation.  Ref.~\GiLione\ proposed that this limit, the  {\it gravitational locality bound}, represents a limit on the domain of validity of local quantum field theory.  In short, the proposal is that the usual axioms of local field theory cease to apply to states of the form \twophi\ for for which \schwloc\ is violated; this is hypothesized to represent a {\it fundamental} limit on the regime in which local field theory can be recovered as an approximation to a more fundamental theory including quantum gravity.\foot{If the tidal excitation of the preceding section contributes to nonlocal dynamics, such a locality bound may be tightened beyond \schwloc\ in string theory.}

Another less classical way of stating this hypothesis is the following. Gravitational scattering amplitudes grow with energy.  When they reach order unity, perturbation theory breaks down.\foot{Specifically, we refer to the breakdown of the eikonalized gravitational amplitudes.}  One can ask what physics unitarizes the theory in this regime.  The hypothesis is that this physics is quantum mechanical, but fundamentally nonlocal.  

One might ask more generally what circumstances are likely to allow recovery of local quantum field theory as a limit.  For example, consider a state created by a collection of $N$ operators,
\eqn\Nstate{\phi_{x_1,p_1}\cdots \phi_{x_N,p_N} |0\rangle\ .}
A natural conjecture is that such a state fails to obey the Wightman axioms in the case where the greatest distance $|x_i-x_j|$ is less than the Schwarzschild radius of the 
combined center-of-mass energy, $R_S(|\sum_i p_i|)$.  This ``generalized locality bound" rests on a quantum analog of the hoop conjecture\refs{\Thorne}.

These bounds state that in regimes where gravity becomes strong enough to form a black hole, local quantum field theory must fail.  At first sight this may sound surprising and even wrong.  For example, a large black hole formed from a collection of many low-energy particles is expected to have a good semiclassical description near its horizon, where curvatures are weak.  Moreover, it has long been believed that semiclassical black holes form in ultra-planckian collisions at sufficiently small impact parameter.  Ref.~\refs{\BaFi} gave one discussion of this, and argued for the validity of the semiclassical description in the high-energy limit.  Further work has solidified this viewpoint:  ref.~\refs{\EaGi} argued (following \refs{\Penrose}) for the existence of a trapped surface in the geometry of a high-energy collision (for further description of its shape in $D>4$, see \refs{\YoNa}), and \refs{\Hsu,\GiRy} gave further justification of the validity of the semiclassical expansion. 

However, while the semiclassical approximation certainly appears to describe the gross features of a black hole with large horizon, the black hole information paradox (and its proposed resolution in favor of unitary evolution) strongly suggests that a detailed quantum description of a black hole does not respect local field theory and is fundamentally nonlocal.  Indeed, one aspect of this idea has been encoded in the statement of the holographic principle, since that says that a black hole has far fewer degrees of freedom than  na\"\i ve field theory would predict.

This suggests that a more correct picture is that the semiclassical approximation serves as a kind of mean field approximation, which summarizes macroscopic features of black hole formation and evaporation in regions away from the singularity.  Moreover, local quantum field theory may be a valid approximate description of certain phenomena, for example experiments conducted, for a time, in the lab of an observer falling into a big black hole.
However, this viewpoint suggests that a semiclassical expansion should {\it not} suffice to give a full quantum description of finer details of black hole evolution.

\subsec{Locality, black holes, and information}

In short, we have argued that strong gravitational effects lead to a breakdown of locality.  With present technology we also can't rule out some violation of locality at even longer distances, perhaps due to tidal string excitation.  A more complete understanding of the role of strings requires improved treatment of high-energy scattering.  And,
a more complete explanation of gravitational nonlocality requires a deeper understanding of the resolution of the information paradox.  't Hooft  and Susskind, as well as \BHMR, suggested that this paradox was likely to be resolved by some form of nonlocality in black hole evaporation, such that information  escapes in the Hawking radiation.  This was pursued and elevated to the level of principle, in the {\it holographic principle} of \refs{\tHooholo\Sussholo\Bous-\Bousrev} and the {\it principle of black hole complementarity} of \refs{\tHooftRE\tHooftFR\tHooftRB-\SusskindIF}.  But, while these hypothesized principles serve as summaries of the expected properties of black holes, consistent with the picture that information escapes in Hawking radiation, they do not resolve the information paradox by explaining precisely what is wrong with Hawking's original argument\refs{\Hawkunc} that black holes destroy information.

Ref.~\refs{\GiLitwo} argued that the loophole in Hawking's derivation of information loss is a faulty assumption:  that one can represent the total Hilbert space in terms of independent Hilbert spaces inside and outside the black hole.  While this decomposition would be implied by local quantum field theory, it could be rendered untrue by the statement that local field theory fails in situations of extreme kinematics, for example parameterized by  the locality bound, \schwloc, or its generalization.  Specifically, ref.~\GiLitwo\ proposed that nonlocality implies a failure of the total Hilbert space to decompose in a controlled approximation in the black hole context. 
This suggestion was amplified in \Ginew: to simultaneously describe information of infalling particles and outgoing late-time Hawking particles, one must compare modes at extremely large relative boosts, which apparently violate the locality bound.\foot{While we are focussing on gravitational effects, nonlocality arising from tidally distorted string effects (or even long strings, if they were somehow relevant) would have the same consequences.}  Thus while a semiclassical treatment like Hawking's is suitable for gross features of black hole dynamics, when one asks finer questions, for example regarding the presence of quantum information in Hawking radiation, one plausibly enters into a domain where the semiclassical approximation fails and local quantum field theory ceases to be a good approximation to physics.

A similar picture relying on suggested nonlocal effects due to long strings was earlier described in \LPSTU, but those authors were not able to agree that they had found a physical effect\refs{\Porev}, and moreover, the discussion of section two failed to reveal such effects but does indicate nonlocal effects due to gravitational mechanisms. This is a satisfying viewpoint:  it suggests that the physics resolving the information paradox can mirror the generality of the physics from which the paradox originated.

The  reader may recognize a certain circularity in the above arguments: a quantum description of black holes based on local physics is not applicable because one must discuss modes for which strong gravity is relevant, and when strong gravity is relevant, locality should break down.  Thus while this picture is self-consistent, and might be thought of as a sort of ``black hole bootstrap," it is not derived from a more fundamental dynamics.  Rather, it stems from the observation that in certain circumstances there is no evidence that local field theory should emerge as a good approximation to a more complete dynamics, together with the perfectly reasonable assumption that this dynamics will not behave like local field theory.  A more complete story would require knowledge of the underlying dynamics.  While many feel that string theory should give a full quantum theory of gravity, string theory is still presently woefully unable to address questions in the regimes of interest, where strong gravitational effects are manifest.  It may even be that a more radical underpinning is needed; for ideas in this direction see the work of Banks and Fischler \refs{\BaFiholo}, and references therein, or \PiazzaGU.

To conclude the discussion of black holes and information it is useful to consider possible connections with other approaches to these problems.  One suggestion for the fate of information falling into a black hole is the ``final state" proposal of Horowitz and Maldacena\refs{\HoMa}, which proposes the existence of a unique final state boundary condition at the singularity.  This would mean that information has been eliminated from an excitation falling into the black hole by the time it reached the singularity; the mechanism for its transmittal to the black hole exterior is less clear.  The proposal based on nonlocal gravitational physics
would suggest a picture similar but different.  Specifically, the above discussion and that of \refs{\GiLitwo,\Ginew} argues that one cannot assume the presence of independent Hilbert spaces inside and outside the black hole, at least when discussing the Hawking radiation.   Indeed, this argument  and its reinforcement in \Ginew\ might ultimately be regarded as a deeper justification for and extension of the principle of black hole complementarity.  If this is the case, but one nonetheless were to attempt to describe the system in terms of independent Hilbert spaces, one might anticipate that the lack of independence manifests itself in dynamics that leads to reduction of the internal part of the Hilbert space to a unique state.  Thus, the Horowitz-Maldacena proposal could be a useful picture to describe the dynamics in the complementary picture appropriate to an outside observer.  On the other hand, if one takes complementarity seriously, there could be a complementary picture that describes the observations of an observer falling into a black hole in terms of a local field theory approximation appropriate to that observer; after all, for a large black hole, such an observer could have a very long lifetime in which local physics is a very good approximation.

It is also worthwhile to comment on the possible relation of the locality bound to statements of the holographic principle, such as the Bousso bound \refs{\Bous,\Bousrev}, which essentially state that the number of degrees of freedom in a region surrounded by area $A$ is proportional to $A$.  Such an assertion in particular implies that local quantum field theory must fail, since for a region of size $R$ local theory would predict a number of states growing as $R^3$.  On the other hand, area bounds do not directly address other circumstances where local field theory should fail, for example the situation where a pure state consisting of two colliding high-energy particles forms a horizon.  Conversely, one might ask whether reasoning based on the locality bound implies area bounds.  In a sense this is close to being true.  If one asks how many degrees of freedom one can excite in a region before the energy is sufficient to form a horizon around that region, in the spirit of our generalized locality bound, a na\"\i ve field-theory estimate is \refs{\tHooholo,\CKN,\GMH} $N\sim R^{3/2}$.  This is off by a power of $R^{1/2}$, possibly arising from gravitational degrees of freedom.  This suggests that a derivation of the holographic principle, stated as an area bound, might be accomplished from a more general statement about limitations on local degrees of freedom with  a more complete understanding of gravitational dynamics.  In this sense, the locality bound in such a setting could be part of a more fundamental explanation of the holographic principle.

\newsec{Conclusion}

The approach of this paper has been to assume that quantum field theory and general relativity are valid until they're forced to fail.  The world could work in other ways, for example, if general relativity is only a long-distance effect and gravity is replaced by some more fundamental dynamics at a scale before it becomes strong.  But the arguments of this paper are more in keeping with the principle of parsimony -- we know gravity and field theory exist, so we simply push them to their limits.

The locality bound then serves as one parameterization of the boundaries of the region of which we are ignorant: the regime where local quantum field theory breaks down and some more fundamental dynamics is relevant. This is useful as it indicates where arguments based on local field theory reasoning can't utilized.  For example,  refs.~\refs{\GiLitwo,\Ginew} argued for the failure of a controlled semiclassical approximation in Hawking's derivation\Hawkunc\ of information loss.  A more complete story may result from consideration of the locality bound, which suggests that this failure is associated with a regime where nonlocal effects are important.  
Related arguments may also suggest general features of what physical effects emerge from beyond the domain of applicability of local field theory, and it would be particularly interesting to investigate what could be said about such effects in other contexts.  

Even more interesting would be to make headway on describing a fundamental underlying dynamics in the terra incognita whose borders are specified by the locality bound.  The expectation has grown that such dynamics is quantum mechanical, but nonlocal, and the picture of this paper is fully consistent with this viewpoint.  String theory may ultimately provide this dynamics, but so far has had little success in the regions of interest.  Or perhaps something more radically nonlocal is needed; for attempts in this direction, see \refs{\BaFiholo,\PiazzaGU}.

The constants of nature $c$ and $\hbar$ have been each associated with revolutions in physics, and a revolution has likewise been predicted to emerge from the understanding of the dynamics associated with Newton's constant $G$.  Arguments such as those discussed here, and their antecedents in the holographic principle,
outline what appears to be a major theme of this revolution:  the new physics of $G$ involves radical reduction of physical degrees of freedom not just on short distance scales, but on {\it all} distance scales.  

So far one can only discern such outlines, based on general reasoning, which, in the case of strong gravitational effects, have been argued to be self-consistent.  In particular, we cannot yet derive such limitations from a complete theoretical framework.  Our situation may be likened to that at the birth of quantum mechanics, described in Bohr's words: ``... in atomic physics the existence of the quantum of action has to be taken as a basic fact that cannot be derived from ordinary mechanical physics."  This existence of the quantum of action of course presents fundamental limits on degrees of freedom and in particular on measurement of complementary variables.  Today, through the evident breakdown of our existing theoretical constructs, specifically general relativity and local quantum field theory, we appear to be seeing
outlines of basic principles of nonlocal physics, associated to $G$, which further limit both degrees of freedom and observations. 
How precisely these principles embed into a deeper theory is  yet to be discovered, and indeed their complete statement may have to be assumed as part of formulating that physics.

\bigskip\bigskip\centerline{{\bf Acknowledgments}}\nobreak

I have benefitted with conversations with T. Banks, D. Gross, S. Hossenfelder, D. Marolf, H. Ooguri, J. Polchinski, A. Strominger, L. Susskind, and G. Veneziano.  This work
was supported in part by Department of Energy under Contract DE-FG02-91ER40618.

\listrefs
\end